 \documentclass[pra, aps, twocolumn, groupedaddress, showpacs, superscriptaddress]{revtex4-2}
 \usepackage{amssymb,amsmath,amsthm,color,graphicx,times,graphicx}\usepackage[caption=false]{subfig}
 \usepackage{hyperref}
 \usepackage{enumerate}
 \usepackage{graphicx}
 \usepackage{dsfont}
 \usepackage{times}
 \usepackage{bbold}
 \usepackage{color}

 \newcommand{\bra}[1]{\langle{#1}|}
 \newcommand{\ket}[1]{|{#1}\rangle}
 
 \usepackage{orcidlink}
 \usepackage{hyperref}
 \hypersetup{
 	colorlinks=true,
 	linkcolor=blue,
 	filecolor=blue,      
 	urlcolor=blue,
 	citecolor=violet,
 	pdfauthor={kkk},
 	pdftitle={},
 	pdfcreator={Abdallah Slaoui},
 }
 
 \providecommand{\openone}{\leavevmode\hbox{\small1\kern-4.3pt\normalsize1}}

 \theoremstyle{plain}

 \theoremstyle{definition}

\begin{document}
 \title{Effects of DM and KSEA interactions on entanglement, Fisher and Wigner-Yanase information correlations of two XYZ-Heisenberg-qubit states under a magnetic field}
 \author{S. Gaidi \orcidlink{0009-0003-9635-0274}}\affiliation{LPHE-Modeling and Simulation, Faculty of Sciences, Mohammed V University in Rabat, Rabat, Morocco.}
 \author{A. Slaoui \orcidlink{0000-0002-5284-3240}}\email{Corresponding author: abdallah.slaoui@um5s.net.ma}\affiliation{LPHE-Modeling and Simulation, Faculty of Sciences, Mohammed V University in Rabat, Rabat, Morocco.}\affiliation{Centre of Physics and Mathematics, CPM, Faculty of Sciences, Mohammed V University in Rabat, Rabat, Morocco.}\affiliation{Center of Excellence in Quantum and Intelligent Computing, Prince Sultan University, 11586, Riyadh, Saudi Arabia.}
 \author{A-B. A. Mohamed}\affiliation{Department of Mathematics, College of Science and Humanities, Prince Sattam bin Abdulaziz University, Saudi Arabia.}\affiliation{Department of Mathematics, Faculty of Science, Assiut University, Assiut, Egypt.}
 \author{M. EL Falaki}\affiliation{Laboratory of Innovation in Science, Technology and Modeling, Faculty of Sciences, Chouaib Doukali University, El Jadida, Morocco.}\affiliation{Centre of Physics and Mathematics, CPM, Faculty of Sciences, Mohammed V University in Rabat, Rabat, Morocco.}
 \author{R. Ahl Laamara \orcidlink{0000-0001-8410-9983}}\affiliation{LPHE-Modeling and Simulation, Faculty of Sciences, Mohammed V University in Rabat, Rabat, Morocco.}\affiliation{Centre of Physics and Mathematics, CPM, Faculty of Sciences, Mohammed V University in Rabat, Rabat, Morocco.}

 \begin{abstract}
 We employ entanglement negativity, local quantum uncertainty (LQU), and local quantum Fisher information (LQFI) to characterize thermal entanglement between two XYZ-Heisenberg-qubit states under the influence of Dzyaloshinsky–Moriya(DM) and Kaplan–Shekhtman–Entin-Wohlman–Aharony (KSEA) interactions, as well as a magnetic field and thermal equilibrium temperature. A comparative examination reveals similar behaviors among these correlation measures. For the antiferromagnetic scenario, we observe that increasing the DM interaction parameter $D_{z}$ enhances thermal entanglement. Conversely, in the ferromagnetic case, the behavior of thermal entanglement differs with varying $D_{z}$. Additionally, employing Kraus operators, we explore the performance of these quantifiers under decoherence. Notably, LQFI exhibits greater robustness than negativity and LQU, even displaying a frozen phenomenon at some time under dephasing effects.\par
 
 \vspace{0.25cm}
 \end{abstract}
 \date{\today}
 	
 	\maketitle
 	\section{Introduction}
  
 Quantum information theory is a rapidly evolving field focused on quantum correlations, including quantum entanglement \cite{Cerf1997,AmgharS2023}. Unlike classical correlations, quantum entanglement enables non-local connections between particles, even over vast distances \cite{Erhard2020}.Recent experiments have demonstrated the presence of quantum entanglement in many-body systems, such as optical lattice setups \cite{Mishmash2009}, quantum cryptography \cite{Gisin2002}, and quantum communication protocols \cite{Guo2019,Slaoui2023}. In addition to entanglement, other forms of non-classical correlations have also emerged as valuable resources for various quantum protocols \cite{RahmanAEJ2023,SlaouiS2018}. These include concepts like measurement-induced nonlocality \cite{Hu2015,Luo2011}, quantum discord \cite{Ollivier2001,Dakic2010}, measurement-induced disturbance \cite{Luo2008} and geometric quantum discord \cite{Luo2010,Paula2013}.\par

In the context of many-body physics, directly calculating the density matrix for large-scale quantum systems is computationally challenging \cite{Laflorencie2016}, recent studies have employed techniques like quantum nondemolition (QND) measurements and adaptive unitaries \cite{Chaudhary}, enabling the deterministic creation of large-scale entangled states. These advancements have applications in quantum metrology \cite{Rossi2020,Abouelkhir2023} and teleportation \cite{Wang2015,Kirdi2023}. Our focus extends beyond negativity entanglement to explore advanced quantum correlation functions, specifically local quantum Fisher information (LQFI) \cite{Kim2018,AbouelkhirS2023} and local quantum uncertainty (LQU) \cite{Girolami2013,SlaouiQIP2018}. Examining such non-classical correlations allows us to understand their strengthening compared to entanglement, as well as their emergence or decay during their evolution in quantum systems.\par
 	
Quantum entanglement is universally acknowledged as a fundamental element in quantum information technology. Its distinctiveness from classical counterparts showcases its power as a unique resource \cite{CerfN1997}. This phenomenon has intrigued physicists since the 1930s, with Einstein, Podolsky, and Rosen sparking a long-standing debate regarding the implications for classical and quantum theories \cite{Einstein1935,C.H.Bennett1992}. The complexity of entanglement makes it generally impossible to alter one part of a well-defined entangled system without considering the other part \cite{M.A.Neilsen2001,Degiovanni2021}. This lack of communication between entangled particles leads to the emergence of non-separable states, violating Bell's inequality \cite{PauL2013,Clauser1969}. On the other hand, LQU and LQFI are emerging as powerful tools for investigating quantum correlations, particularly in multipartite systems. This is primarily due to the relative ease of calculating both of these quantities. In fact, LQU utilizes the skew information formalism, a mathematical framework for characterizing measurement uncertainty \cite{E.P.Wigner1963}. Notably, for bipartite qubit-qudit systems, the LQU minimization can be solved analytically. Furthermore, LQU is intricately linked to Fisher's concept of quantum information \cite{S.Luo2003}, making it a crucial component in quantum metrology protocols.\par
 	
Heisenberg spin models have recently become a focal point of research in quantum systems due to their ability to exhibit non-classical correlations. These models are crucial for simulating various systems, including nearest-neighbor interactions between quantum spins and optical lattices \cite{Kloeffel2013}. Researchers have investigated a wide range of properties in Heisenberg models, including negativity \cite{M.Horodecki1996}, heat capacity \cite{T.Mamtimin2013}, and specific cases with Dzyaloshinskii-Moriya (DM) interaction \cite{FedorovaV2022,Ali2024}. Thermal entanglement characteristics in the Heisenberg model have been thoroughly investigated to analyze correlation properties and their potential applications in quantum information \cite{G.F.Zhang2005,Park2019}. Additionally, thermal spin entanglement under an external magnetic field has been a focal point of numerous investigations \cite{M.C.Amesen2001}. Entanglement has been explored using Heisenberg models with DM interaction \cite{X.G.Wang2002,FedorovaV2022}. Several significant works have contributed to exploring quantum correlations in quantum systems described by Heisenberg XX models \cite{X.G.Wang2001,G.F.Zhang2005}. Furthermore, the impact of the DM interaction on quantum correlations in the Heisenberg-XYZ model was explored without considering magnetic field effects.\par
 	
Motivated by recent findings on the influence of DM and KSEA interactions, we investigate their combined impact within a fully anisotropic Heisenberg Model for two spin-$\frac{1}{2}$ particles.  Our work focuses on a detailed analysis of three key measures of quantum correlations: negativity, local quantum uncertainty, and local quantum Fisher information. We consider the model in thermal equilibrium.  The paper is structured as follows. Section~\ref{sec2} introduces the Hamiltonian model and correlation quantifiers. Section~\ref{sec3} presents the numerical results for entanglement properties in the model and discusses the behavior of quantum correlations. Section~\ref{sec4} examines the dynamics of the quantum correlation quantifiers. Finally, conclusions are provided in Section~\ref{sec5}.
 	
 	\section{Theoretical Model}\label{sec2}
 	Heisenberg spin chains are good candidates for describing non-classical correlations in quantum mechanics. Compared to other systems, they exhibit unique characteristics, such as long-range entanglement and integrability under specific conditions \cite{FuHC2007,BayatA2008}. In this study, we present a model of a two-qubit Heisenberg XYZ chain with Dzyaloshinskii-Moriya (DM) and KSEA interactions under an external magnetic field. The system Hamiltonian is characterized by
 	\begin{align}
 			\mathcal{H}=&J_x\sigma_{1}^x\sigma_{2}^x+J_y\sigma_{1}^y\sigma_{2}^y+J_z\sigma_{1}^z\sigma_{2}^z+D_z(\sigma_{1}^x\sigma_{2}^y-\sigma_{1}^y\sigma_{2}^x)\notag\\
 			&+ \Gamma_z(\sigma_{1}^x\sigma_{2}^y-\sigma_{1}^y\sigma_{2}^x)+B(\sigma_{1}^z+\sigma_{2}^z)
 	\end{align}
Here, the coupling constants are denoted as $J_x$, $J_y$, and $J_z$, while $\sigma_{i}$, where $i$ ranges from $1$ to $2$, represents the Pauli matrices.  In the computational basis comprising states ${\ket{00},\ket{01},\ket{10},\ket{11}}$, the Hamiltonian matrix is structured as follows
 		\begin{align}
 			\mathcal{H}&= \left(2B+J_{z}\right)\ket{00}\bra{00}+\left(J_z-2B\right)\ket{11}\bra{11}\notag\\&-J_{z}\left[\ket{01}\bra{01}+\ket{10}\bra{10}\right] +\left(J_{x}+J_{y}\right)\left[\ket{10}\bra{01}+\ket{01}\bra{10}\right]\notag\\&+\left(J_{x}-J_{y}\right)\left[\ket{11}\bra{00}+\ket{00}\bra{11}\right] +2iD_{z}\left[\ket{10}\bra{01}-\ket{01}\bra{10}\right]\notag\\&+2i\Gamma_{z}\left[\ket{00}\bra{11}-\ket{11}\bra{00}\right].\label{H}
 		\end{align}
 	The eigenvalues of the Hamiltonian (Eq.(\ref{H})) are given by
 	\begin{equation}
E_{1,2}=-J_z\pm m_{1}, \hspace{1cm}E_{3,4}=J_z\pm m_{2}
 	\end{equation}
 where
 \begin{align}
&m_{1}=\sqrt{4 D_z^2+(J_x+J_y)^2}, \notag\\ &m_{2}=\sqrt{4 B^2+4 \Gamma_z^{2}+(J_x-J_y)^2}.
\end{align}
To investigate the thermal effects on quantum correlations, we consider the model in thermal equilibrium, described by the Gibbs canonical ensemble for thermal equilibrium. This allows us to express the system's thermal state as
 	\begin{equation}
 		\rho =\frac{1}{Z}\exp\left[-\beta  \mathcal{H}\right],
 	\end{equation} 
 	with the partition function, denoted as $Z = \sum_{i} e^{-\beta E_i}$, where $E_{i}$ signifies the energy of state $i$ and $\beta = \frac{1}{k_{B}T}$ relates the inverse temperature to the Boltzmann constant $k_{B}$. In the customary natural unit system where $\hbar=k_{B}=1$ (these constants are set to $1$), the density matrix of the system is given by
 	\begin{equation}
 		\begin{aligned}
 			\rho=
 			&\rho_{11} \ket{00}\bra{00} + \rho_{14} \ket{11}\bra{00} + \rho_{23} \ket{10}\bra{01} \\
 			&+ \rho_{22} \ket{01}\bra{01} + \rho_{23}^*\ket{01}\bra{10} +\rho_{33}\ket{10}\bra{10}\\ 
 			&+ \rho_{14}^* \ket{00}\bra{11} + \rho_{44} \ket{11}\bra{11},
 		\end{aligned}\label{Eq6}
 	\end{equation}
and their elements are expressed as
 	\begin{align}
 		&\rho_{11}	=\frac{e^{-\beta J_z} \left(\cosh (\beta r_3)-\frac{2 B \sinh (\beta r_3)}{r_3}\right)}{Z}, \\
 		&\rho_{14}= \frac{e^{-\beta J_z} \sinh (\beta r_3) (2 i \Gamma_z - J_x+ J_y)}{r_3 Z},\\
 		&\rho_{22}=\rho_{33} = \frac{e^{\beta J_z} \cosh (\beta r_2)}{Z}, \\
 		&\rho_{23}= \frac{e^{\beta  J_z} (-( J_x+ J_y) \sinh (\beta r_2)+2 i D_z \cosh (\beta  r_2))}{r_2 Z},\\
 		&\rho_{44}=  \frac{e^{-\beta J_z} \left(\frac{2 B \sinh (\beta r_3)}{r_3}+\cosh (\beta r_3)\right)}{Z},
 	\end{align}
 	with
 	\begin{align}
 		r_{1}=&\sqrt{4 \Gamma_z^2+(J_x-J_y)^2},\hspace{1cm}
 		r_{2}=\sqrt{4  D_z^2+(J_x+J_y)^2},\notag
 		\\
 		&r_{3}=\sqrt{4 \Gamma_z^2 +4 B^2+(J_x-J_y)^2}.
 	\end{align}
 	Local unitary transformations acting on each qubit can be applied to remove the phase factors from the off-diagonal elements of the density matrix \cite{slaoui2019}
 	\begin{equation}
\left|0\right\rangle_{\lambda}\rightarrow \exp\left[\frac{i}{2}\left(\varphi_{14}+(-1)^{\lambda} \varphi_{23}\right) \right]\left|0\right\rangle_{\lambda}, \hspace{0.5cm}\lambda=1,2
 	\end{equation}
 	where $\rho_{14}/|\rho_{14}|=\exp\left(i\varphi_{14}\right)$ and $\rho_{23}/|\rho_{23}|=\exp\left(i\varphi_{23}\right)$, with the density matrix $\rho$ preserving its rank and having positive off-diagonal entries. In fact, local unitary transformations do not affect the physical properties of a quantum system. Removing unnecessary phase factors simplifies calculations and analysis without losing essential information. This is particularly useful for mixed states and in studying non-classical correlation properties. The state (\ref{Eq6}) can be expressed as
  \begin{equation}
 		\begin{aligned} 
 			\rho \rightarrow \varrho=&
 			\hat{\varrho}_{11} \ket{00}\bra{00} + \hat{\varrho}_{14} [\ket{00}\bra{11}+ \ket{11}\bra{00}] \\
 			&+ \hat{\varrho}_{22} \ket{01}\bra{01} + \hat{\varrho}_{23} [\ket{01}\bra{10}+ \ket{10}\bra{01} ]  \\
 			& +\hat{\varrho}_{33} \ket{10}\bra{10}  + \hat{\varrho}_{44} \ket{11}\bra{11},
 		\end{aligned}\label{eq14}
 	\end{equation}
 	with
  \begin{equation}
      \hat{\varrho}_{ii}=\rho_{ii}, \hspace{0.5cm}{\rm where} \hspace{0.5cm} i=1,2,3,4,
  \end{equation}
 	\begin{equation}
\hat{\varrho}_{14}=|\rho_{14} |= \frac{r_1 e^{-\beta j_z} \sinh (\beta r_3)}{r_{3} Z},
\end{equation}
\begin{align}
 		&\hat{\varrho}_{23}=|\rho_{23}|=\frac{e^{\beta J_z} \sqrt{4D_z^2 \cosh ^2(\beta r_{2})+(J_{x}+J_{y})^2 \sinh ^2(\beta  r_{2})}}{r_2 Z}.
 	\end{align}

The solutions to the equation (\ref{eq14}) are its eigenvalues, which are given by
 	\begin{equation}
\eta_1 =\frac{e^{\beta J_z}}{Z} (\cosh (\beta r_{2})-\frac{ \xi}{r_{2}}),\hspace{1cm}\eta_{3}= \frac{e^{-\beta(r_{3} +J_{z} )}}{Z},
 	\end{equation}
 	\begin{equation}
\eta_{2}=\frac{e^{\beta J_{z}}}{Z} (\cosh (\beta r_{2}) +\frac{\xi}{r_{2} }),\hspace{1cm}\eta_{4}=\frac{e^{-\beta(r_{3}-J_{z})}}{Z},
 	\end{equation}
 	with
 	\begin{equation}
\xi =\sqrt{4D_{z}^{2}-(J_{x}+ J_{y})^{2} +r_{2}^{2} \cosh (2\beta r_{2})}.
 	\end{equation}
 
 	\section{Dynamics of Non-local Correlations}
 	Here, we consider nonlocal correlations within the framework of a two-qubit XYZ-DM Heisenberg model while considering the influence exerted by KSEA interactions. We will quantify this nonlocality using these three specific quantifiers:
  
 \paragraph*{\bf Entanglement Negativity:} While entanglement entropy effectively quantifies entanglement in pure bipartite states, it becomes inapplicable for mixed states or systems with more than two qubits. Logarithmic negativity ($\mathcal{N}_{L}\left(\rho\right)$), introduced by Vidal and Werner \cite{Peres1996,Vidal2002}, offers a valuable alternative. Notably, $\mathcal{N}_{L}\left(\rho\right)$ is easily computable for any mixed state of a bipartite system and is based on the concept of negativity ($\mathcal{N}\left(\rho\right)$) \cite{Miranowicz2004}. Negativity refers to the absolute sum of negative eigenvalues obtained by partially transposing the density matrix with respect to first subsystem $A$, \cite{M.B.Plenio2005}, i.e. $\rho^{T_{A}}$;
 	\begin{equation}
 		\mathcal{N}\left(\rho\right)= \frac{||\rho^{T_{A}}||_{1}-1}{ 2}=\sum_{i}\mu_{i},
 	\end{equation}
 where $\mu_i$ represent the negative eigenvalue of the partially transposed density matrix $\rho^{T_{A}}$ (where $T_{A}$ indicates the partial transpose with respect to subsystem $A$), and the trace norm $||\rho^{T_{A}}||_{1}={\rm Tr}\left(\sqrt{\rho^{T_{A}}\rho^{T_{A}^{\dagger}}}\right)$. The negativity for a two-qubit state is then given by
 	\begin{equation}
 		\mathcal{N}\left(\rho\right)= \max\{0,-2\mu_{\min}\},\label{eq21}
 	\end{equation}
 	where this measure is non-zero only for entangled states, making it a valuable tool to quantify the degree of entanglement in any mixed bipartite state. The negative eigenvalues of the partially transposed density matrix  for the state (\ref{eq14}), in the computational basis, take the form

 	\begin{equation}
 		\begin{aligned} 
 			\varrho^{T_{1}}=& \hat{\varrho}_{11}\ket{00}\bra{00}+ \hat{\varrho}_{22}\left[\ket{01}\bra{01}+\ket{10}\bra{10}\right]+\\
 			&\hat{\varrho}_{44} \ket{11}\bra{11} + \hat{\varrho}_{23} \left[\ket{11}\bra{00}+ \ket{00}\bra{11}\right]\\
 			&+ \hat{\varrho}_{14}\left[\ket{10}\bra{01}+ \ket{01}\bra{10}\right],
 		\end{aligned}
 	\end{equation}
 	and its eigenvalues are given by 
 	\begin{align}
 		e_1&=\frac{e^{-\beta J_z}}{Z} \left(\cosh (\beta r_3)-\frac{\sqrt{e^{4 \beta J_z}r_3^2\chi +4B^2r_2^2 \sinh ^2(\beta r_3)}}{r_2 r_3 }\right),\\
 		e_2&=\frac{e^{-\beta J_z}}{Z} \left(\cosh (\beta r_3)+\frac{\sqrt{e^{4 \beta J_z}r_3^2\chi +4B^2r_2^2 \sinh ^2(\beta r_3)}}{r_2 r_3 }\right),\\
 		e_3&=\frac{e^{\beta J_z} \cosh (\beta  {r_2})}{Z}-\frac{e^{-\beta {j_z}} r_1 \sinh (\beta r_3)}{ {r_3} Z},\\
 		e_4&=\frac{e^{\beta J_z} \cosh (\beta  {r_2})}{Z}+\frac{e^{-\beta {j_z}} r_1 \sinh (\beta r_3)}{ {r_3} Z},
 	\end{align}
 	where
 	\begin{align*}
 		\chi= \sqrt{ 4 D_z^2\cosh^2( \beta r_2) +(j_x+j_y)^2 \sinh^2 \beta( r_2 )}.
 	\end{align*}
 	Hence, the negativity is described by equation (\ref{eq21}) with
 	$ \mu_{min}= \min\{e_1,e_2,e_3,e_4\}$.\\
  
\paragraph*{\bf Local quantum uncertainty:} LQU, introduced by Girolami et al.\cite{GiloramiD2013}, offers an alternative measure of quantum correlations. Notably, LQU is analytically computable for any generic quantum state. It builds upon the concept of skew information, introduced by Wigner and Yanase in 1963 \cite{Wigner1963}, which was originally introduced to quantify information content in mixed states. Skew information plays a crucial role in the quantifying measurement uncertainty of observables \cite{Luo2005}, relating to Fisher information for quantum metrology \cite{SlaouiB2019}, and distinguishing quantum states \cite{Luo2012}. Additionally, skew information serves as a quantifier of quantum coherence \cite{Du2015,Girolami2014}. Actually, LQU carries significant implications across various domains of quantum information science. It is defined as the minimum skew information associated with a single measurement observable on a subsystem \cite{AhmadN2022}. This foundational concept not only sheds light on the fundamental limits of measurement precision within quantum systems but also lays the groundwork for the development of robust quantum communication protocols. Furthermore, exploring local quantum uncertainty opens avenues for advancing quantum computing technologies, providing crucial insights into the intricate interplay between quantum measurements and computational tasks. The analytical expression of LQU for any qubit-qudit quantum system can be given by
 	\begin{equation}
 		{\cal LQU}\left(\varrho \right)=1-\max\{\epsilon_{1},\epsilon_{2},\epsilon_{3}\}\label{eq27}
 	\end{equation}
 	Where $\epsilon_{i}$ ($i=1,2,3$) refers to  the eigenvalues of  the 3 ×3 symmetric matrix ${\cal W}$ whose entries are
 	\begin{equation}
 		\left( {\cal W}\right)_{ij}=Tr \{ \sqrt{\rho}\left( \sigma_{i}\otimes I\right) \sqrt{\rho}\left( \sigma_{j}\otimes I\right)\},\label{Eq28}
 	\end{equation}
 	with ($i,j=x,y,z$) and $\sigma_{i,j}$ represent the Pauli matrices. Within the context of the model we are considering, the elements of the matrix ${\cal W}$ (Eq.(\ref{Eq28})) are given by
 	 \begin{align}
 	{\cal W}_{11}= \chi_1 +\frac{16r_1 \sqrt{\Omega} \sinh[\beta r_3]}{4\chi_1 r_2 r_3 Z^2},
 	\end{align}
 	 \begin{align}
 		{\cal W}_{22} &= \chi_1 -\frac{16r_1 \sqrt{\Omega} \sinh[\beta r_3]}{4\chi_1 r_2 r_3 Z^2},
 		\end{align}
 	\begin{equation}
 		{\cal W}_{33} = \frac{1}{2} \chi_4^2 -\frac{16 e^{2\beta J_z}\Omega}{8 r_2^2 Z^2 \chi_3^2}-\frac{16 e^{-2 \beta J_z}( r_1^2-4 B^2 )\sinh ^2(\beta r_3)}{8 r_3^2 Z^2 \chi_2^2},
 	\end{equation}
 	with 
 	\begin{equation}
\Omega=4D_z^2 \cosh[\beta r_2]^2 +(J_x+J_y)^2\sinh[\beta r_2]^2
 	\end{equation}
 	and $\chi_i$ are given respectively by
 	\begin{align}
	\chi_1=\frac{1}{Z}&\left(\frac{1}{\sqrt{2r_2}}(\sqrt{e^{-\beta r_3}}\tau_1^{-}+\sqrt{e^{\beta r_3}} \tau_1^{+})\right. \notag\\&\left.  +(J_x+J_y)\frac{\sqrt{2}e^{\beta J_z}}{r_2}+e^{-\beta J_z}\right),
 	\end{align}
 	\begin{align}
&\chi_{2}= \frac{1}{\sqrt{Z}}\left(\tau_2^{-}+\sqrt{e^{-\beta(J_z-r_3) }} \right),\\&
\chi_{3}= \frac{1}{\sqrt{Z}}\left( \tau_2^{+}+\sqrt{e^{-\beta(J_z+r_3) }}\right),
 	\end{align}
 	\begin{align}
 		\chi_{4}= \frac{1}{4\sqrt{Z}}&\left( \left(2\sqrt{\beta(-J_z+r_3)} +\sqrt{2}\tau_3^{-}\right)^2 \right. \notag\\&\left. +\left(2\sqrt{-\beta(J_z+r_3)} +\sqrt{2}\tau_3^{+}\right)^2 \right),
 	\end{align}
 	\begin{align}
\tau_1^{\pm}&=\frac{\sqrt{ \left(2  r_2 \cosh (\beta r_2) \pm \sqrt{\upsilon_{+}}\right)}}{\sqrt{2r_{2} }},
 	\end{align}
 	\begin{align}
 			\tau_3^{\pm}&=\sqrt{\frac{e^{\beta  J_z}(2r_2\cosh (\beta r_2) \pm \sqrt{\upsilon_{-}})}{r_2} },
 	\end{align}
        \begin{align}
               \tau_{2}^{\pm} &= \sqrt{\frac{e^{\beta  J_z} {\cal y}}{\sqrt{2} r_2 }+ e^{\beta J_z} \cosh (\beta r_2)}
        \end{align}
 where
 	\begin{equation}
\upsilon_{\pm}=8 D_z^2-2 (J_x+J_y)^{2}+2r2^{2} \cosh (2 \beta r_2),
 	\end{equation}
  and
  \begin{equation}
      {\cal Y}=\sqrt{r_2^2 \cosh (2 \beta r_2)+4 d^2-(J_x+J_y)^2}.
  \end{equation}
 Hence, the non-classical correlation through LQU is equal to
 	\begin{equation}
 		{\cal LQU}\left(\varrho \right)=1-\max\{{\cal W}_{11},{\cal W}_{22},{\cal W}_{33}\}.
 	\end{equation}
 	
 \paragraph*{\bf Local quantum Fisher information $\mathcal{Q}$:} Born from quantum metrology, Fisher information quantifies the precision achievable in measurements within the quantum domain. In estimation theory, it reveals the amount of information a parameter contains in measurement results. Higher Fisher information indicates greater ease in distinguishing a quantum state from similar states for that parameter. Consequently, it serves as the ultimate benchmark for evaluating the accuracy of parameter estimation protocols \cite{slaoui2019,Helstrom1976}. Recent efforts have been devoted to evaluating the QFI trajectory to determine the role of quantum correlation in metrology. An elucidated study demonstrated that quantum entanglement results in notable improvements in both the efficiency and precision of parameter estimation \cite{Chapeau-Blondeau2017,Chapeau-Blondeau2016}. Analogous to entanglement, the $\mathcal{Q}$ is a genuine measure of non-classical correlations, aiding in understanding the role of quantum correlations in the phase estimation protocol. For a parametric state $\rho_\theta $ contingent upon $\theta$, the QFI is defined as \cite{Luo2009}
 	\begin{equation}
 		\mathcal{Q}(\rho_\theta)=\frac{1}{4}{\rm Tr}[\rho_\theta L_\theta^2],
 	\end{equation}   
 	where $L_{\theta}$ is the symmetric logarithmic derivative operator satisfying
 	\begin{equation}
 		\frac{\partial \rho_\theta}{\partial \theta}=\frac{1}{2}\left(L_\theta \rho_\theta+\rho_\theta L_\theta\right).
 	\end{equation}
 	and by using the unitary transformation $U_\theta=e^{i\theta}$, we obtain the parametric state $\rho_\theta$ generated by a Hermitian operator $H$, i.e. $\rho_\theta=U_\theta\rho^{\dagger}U_\theta$, then the QFI can be evaluated as
 	\begin{equation}
 		\mathcal{Q}(\rho, H)=\frac{1}{2} \sum_{m \neq n} \frac{\left(\lambda_m-\lambda_n\right)^2}{\lambda_m+\lambda_n}|\langle m|H| n\rangle|^{2}.
 	\end{equation}
 	$\mathcal{Q}$ capturing the maximum achievable information gain on a parameter. This scenario takes into account all possible local Hamiltonians that affect only a specific subsystem $A$. The local Hamiltonian's general form is represented as $H_A=\sigma \cdot \mathbf{r}$, where $|\mathbf{r}|=1$ and $\sigma=\left(\sigma_1, \sigma_2, \sigma_3\right)$ \cite{Bera2014,slaoui2019}. Therefore, $\mathcal{Q}$ can be written as
 	\begin{equation}
 		\mathcal{Q}(\rho, H_A)= Tr(\rho H_A^2)-  \sum_{m \neq n} \frac{ 2\lambda_m\lambda_n}{\lambda_m+\lambda_n}|\langle m|H_A| n\rangle|^{2},
 	\end{equation}
 and for any qubit-qudit systems, it is explicitly given as
 	\begin{equation}
 		\mathcal{Q}\left(\rho\right)=1-\lambda_{max}\left({\cal M} \right)\label{eq47}
 	\end{equation}
 	Where $\lambda_{max}$ represents the  largest eigenvalue of the real symmetric $3\times3$ matrix ${\cal M}$, which comprises entries
 	\begin{equation}
 		{\cal M}_{ij}=\sum_{m \neq n} \frac{2 \lambda_m \lambda_n}{\lambda_m+\lambda_n}\left\langle m\left|\sigma_i \otimes I\right| n\right\rangle\left\langle n\left|\sigma_j \otimes I\right| m\right\rangle.
 	\end{equation}
 	After some simplifications, the elements of matrix ${\cal M}$ for the considered model (see Eq.(\ref{eq14})) are given by
 	\begin{equation}
 		{\cal M}=\left(
 		\begin{array}{ccc}
 			{\cal M}_{xx} & {\cal M}_{xy} & 0 \\
 			{\cal M}_{yx} & {\cal M}_{yy} & 0 \\
 			0 & 0 & {\cal M}_{zz} \\
 		\end{array}
 		\right),
 	\end{equation}  
 	and its elements are given by
 	\begin{align}
{\cal M}_{xx}=\frac{2}{r_3}&\left((r_1+r_3)\frac{\eta_1 \eta_3 }{\eta_1+\eta_3}+\frac{\eta_2 \eta_4 }{\eta_2+\eta_4}\right. \notag\\&\left. +(r_3-r_1)\frac{\eta_2 \eta_3 }{\eta_2+\eta_3}+\frac{\eta_1 \eta_4 }{\eta_1+\eta_4}\right),
 	\end{align}
 	\begin{align}
{\cal M}_{xy}&=\frac{-4 i B}{r_3} \left(\frac{\eta_1 \eta_3}{\eta_1+\eta_3}+\frac{\eta_1 \eta_4  }{\eta_1+\eta_4}+\frac{\eta_2 \eta_3  }{\eta_2+\eta_3}+\frac{\eta_2 \eta_4 }{\eta_2+\eta_4}\right),
 	\end{align}
 	\begin{align}
	{\cal M}_{yx}&=\frac{4 i B}{r_3} \left(\frac{\eta_1 \eta_3}{\eta_1+\eta_3}-\frac{\eta_1 \eta_4  }{\eta_1+\eta_4}+\frac{\eta_2 \eta_3 }{\eta_2+\eta_3}-\frac{\eta_2 \eta_4 }{\eta_2+\eta_4}\right),
 	\end{align}
 	\begin{align}
{\cal M}_{yy}=\frac{2}{r_3}&\left((r_3-r_1)\frac{\eta_1 \eta_3 }{\eta_1+\eta_3}+\frac{\eta_2 \eta_4 }{\eta_2+\eta_4}\right. \notag\\&\left. +(r_3+r_1)\frac{\eta_2 \eta_3 }{\eta_2+\eta_3}+\frac{\eta_1 \eta_4 }{\eta_1+\eta_4}\right),
 	\end{align}
 \begin{align}
 {\cal M}_{zz}&=\frac{4 \eta_{1} \eta_2}{\eta_1+\eta_2}+\frac{4\eta_3 \eta_4 r_1^2}{\eta_3+\eta_4 (r_{3}^{4}-4B^2r_3^2)}.  
 \end{align}
 It can be easily shown that the eigenvalues of the matrix ${\cal M}$ defined by equation (37) are equal to: $\lambda_{3}= {\cal M}_{zz}$,
 	\begin{align*}
\lambda_{1}=&\frac{1}{2}\left[{\cal M}_{xx}+{\cal M}_{yy}\right. \notag\\&\left. -\sqrt{{\cal M}_{xx}^2-2 {\cal M}_{xx} {\cal M}_{yy}+4 {\cal M}_{xy} {\cal M}_{yx}+{\cal M}_{yy}^2}\right],
 	\end{align*}
 	\begin{align*}
 	\lambda_{2}=&\frac{1}{2}\left[{\cal M}_{xx}+{\cal M}_{yy}\right. \notag\\&\left. +\sqrt{{\cal M}_{xx}^2-2 {\cal M}_{xx} {\cal M}_{yy}+4 {\cal M}_{xy} {\cal M}_{yx}+{\cal M}_{yy}^2}\right].
 	\end{align*}
 	From the above expressions it is clear that $\lambda_{2}\geq\lambda_{1}$, so the $\mathcal{Q}$ formula is
 	\begin{equation}
\mathcal{Q}\left(\rho\right)=1-\max\left\lbrace\lambda_{2},\lambda_{3} \right\rbrace.
 	\end{equation}
 	
 \section{Numerical results }\label{sec3}
 In this section, we numerically plot nonlocal correlation quantifiers in a two-qubit anisotropic Heisenberg XYZ chain with DM and KSEA interactions, considering the presence of an external magnetic field $B$ across various temperature values. So far, interesting results regarding entanglement in similar systems have been presented in previous works, specifically in Ref. \cite{FedorovaV2022}. To gain a better understanding of the behaviors of the quantities introduced in the last section, we present some figures using Eqs.(\ref{eq21}), (\ref{eq27}), and (\ref{eq47}). Initially, we evaluate these three measures—negativity, LQU, and LQFI—as functions of the $Z$-axis DM interaction for both ferromagnetic and antiferromagnetic exchange couplings. Subsequently, we analyze them as functions of the magnetic field $B$ for a given temperature $T$. Finally, we depict these three measures as functions of the $Z$-axis DM interaction for the antiferromagnetic case ($J_z=2$) at a given magnetic field $B$.\par
 
In Figure (\ref{F1}), our graphical representation highlights the interplay between negativity N, LQU, and LQFI concerning the $Z$-axis DM interaction under varying temperatures T for the antiferromagnetic case $J_z=2$. A notable observation is the symmetrical behavior of quantum correlations around  $D_z = 0$ which points to the inherent symmetry of the DM interaction's impact on the spin system. Physically, the DM interaction introduces anisotropic spin couplings. This balanced behavior around $D_z = 0$  suggests that both positive and negative values of $D_z = 0$ contribute equally to modifying the spin alignment in opposite directions. For small values of \( D_z \), all three quantifiers (negativity, LQU, and LQFI) peak, indicating maximum quantum entanglement. This peak corresponds to the optimal enhancement of quantum correlations by the DM interaction, As the DM interaction influences the system's spin orientations, the correlation decreases leading to a monotonic decline in quantum correlations.
On the other hand, for positive \( D_z \), quantum correlations increase with growing \( D_z \). This rise is due to the DM interaction favoring further non-collinear spin configurations, promoting coherence and, consequently, entanglement. The observed behavior for both positive and negative values of \( D_z \) underscores the role of DM interaction in controlling spin alignment, which directly impacts quantum correlations.
For the ferromagnetic case (\( J_z = -2 \)), a similar initial response to \( D_z \) is observed. However, after the initial decline, quantum correlations begin to increase again. This non-monotonic behavior in the ferromagnetic regime can be attributed to the competition between the ferromagnetic coupling, which favors parallel spin alignment, and the DM interaction, which disrupts this alignment. At higher values of \( D_z \), this  leads to frustration in the spin configurations, which paradoxically enhances quantum entanglement by promoting a more complex and entangled spin state.\par

On the other hand, in Fig.(\ref{F22}), we plot the same quantifiers versus the magnetic field \( B \) for various temperatures in the antiferromagnetic case (i.e., \( J_z = 2 \)). The data illustrate how quantum correlations are affected by both the magnetic field and temperature. As the magnetic field \( B \) varies, quantum correlations display a pronounced temperature-dependent behavior. At lower temperatures, the system remains in a relatively ordered state with limited thermal excitations, which restricts the quantum correlations. As the temperature increases, thermal excitations become more significant, allowing the system to explore a broader range of quantum states and thereby enhancing the correlations. The magnetic field \(B\) modulates the interaction between the qubits, influencing the quantum correlations in a temperature-dependent manner. When the system reaches a high enough temperature, the effects of thermal noise dominate, leading to a plateau in quantum correlations. This plateau signifies a balance between the increasing thermal fluctuations and the intrinsic quantum properties of the system, while the magnetic field \(B\) continues to affect the overall strength and nature of these correlations. The observed trend emphasizes the combined influence of temperature and magnetic field on quantum coherence, illustrating how these factors interplay to shape the system's quantum correlations.\par

In Figure (\ref{F3}), we depict these quantifiers as a function of the DM interaction parameter $D_z$ for the case where $(J_z=-2)$. The curves are obtained at different values of $B$, enabling us to observe the evolution of quantum correlations with the DM interaction parameter $D_z$. It becomes apparent that in the presence of $(J_z=-2)$, an increase in $D_z$ leads to an augmentation in entanglement. Moreover, as $D_z$ reaches very high values, the entanglement tends to remain constant even with further increases in $D_z$.\par

Varying the \(J_{x}\) and \(J_{y}\) coefficients, along with \(J_{z}\), would indeed impact both the LQU and LQFI. Our findings suggest that LQU, which is sensitive to the anisotropic nature of the interactions, would be similarly influenced by changes in \(J_{x}\) and \(J_{y}\). This is consistent with the results reported by Modak et al.\cite{Modak2014}, who observed that varying interaction parameters significantly affects LQU in similar systems. Similarly, LQFI, which measures the system's sensitivity to local perturbations, also varies with different \(J_{x}\) and \(J_{y}\) values, consistent with Tóth’s observations \cite{Toth2005} that LQFI is highly dependent on interaction parameters in spin systems. These results reinforce the importance of considering anisotropic interactions in quantum information measures and underscore the relevance of exploring \(J_{x}\) and \(J_{y}\) in our study to provide a more comprehensive understanding of the system's behavior.\par
 
To analyze the influence of the external magnetic field and temperature on the three different measures in the antiferromagnetic exchange coupling $J_z=2$, we observe their behaviors in Fig.(\ref{F22}). Both the magnetic field and temperature affect the amount of quantum correlation, with an increase in temperature leading to a decrease in quantum correlations. Despite this general trend, the three quantifiers (Fig. (\ref{F22}) a, b, and c) exhibit similar behaviors. Finally, it can be concluded that nonlocal correlations is more enhanced in the case of LQFI, as illustrated in the plots. Next, the behavior of entanglement in Fig.(\ref{F3}) confirms the results obtained previously in Fig.(\ref{F1}). This time, we aim to analyze the effect of the DM interaction parameter $D_z$ on these three measures in the case of antiferromagnetic exchange coupling $J_z=2$. We observe that the variations of $\mathcal{N}$, LQU, and LQFI are consistent as temperature increases: the amount of quantum correlations increases symmetrically for all three cases (Fig.\ref{F1}(a), (b), and (c)). Furthermore, with an increase in $D_z$, the quantum correlation increases and reaches a maximum until it stabilizes at $\mathcal{N}=0.5$ for negativity and $LQU=\mathcal{Q}=1$ for LQU and LQFI.\par

Based on the above discussion, we conclude that negativity, LQU, and LQFI are suitable candidates for measuring nonlocal correlations. Moreover, LQFI appears to be more accurate. We also find that LQFI is a robust method for detecting non-classical correlations. This conclusion aligns with the inequality established in Ref.\cite{slaoui2019}, which demonstrates that LQFI is always greater than LQU.
 	
 	\begin{widetext}
 	
 		\begin{figure}[hbtp]
 			{{\begin{minipage}[b]{.33\linewidth}
 						\centering
 						\includegraphics[scale=0.45]{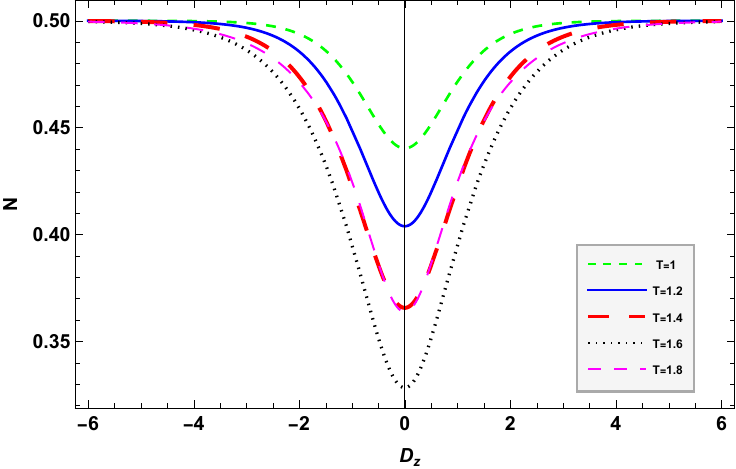} \vfill $\left(a\right)$
 					\end{minipage} \hfill
 					\begin{minipage}[b]{.33\linewidth}
 						\centering
 						\includegraphics[scale=0.45]{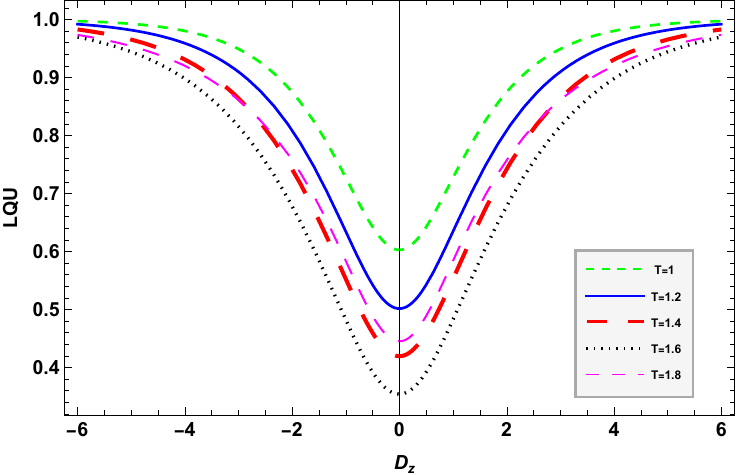} \vfill  $\left(b\right)$
 					\end{minipage} \hfill
 					\begin{minipage}[b]{.33\linewidth}
 						\centering
 						\includegraphics[scale=0.45]{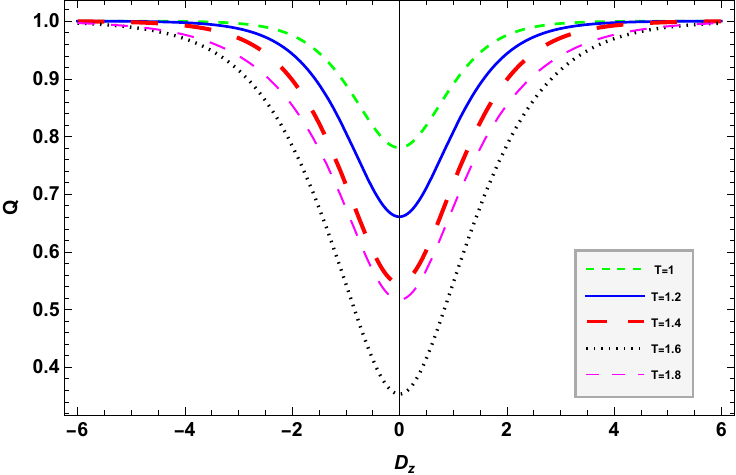} \vfill  $\left(c\right)$
 			\end{minipage}}}

 			{{\begin{minipage}[b]{.33\linewidth}
 						\centering
 						\includegraphics[scale=0.45]{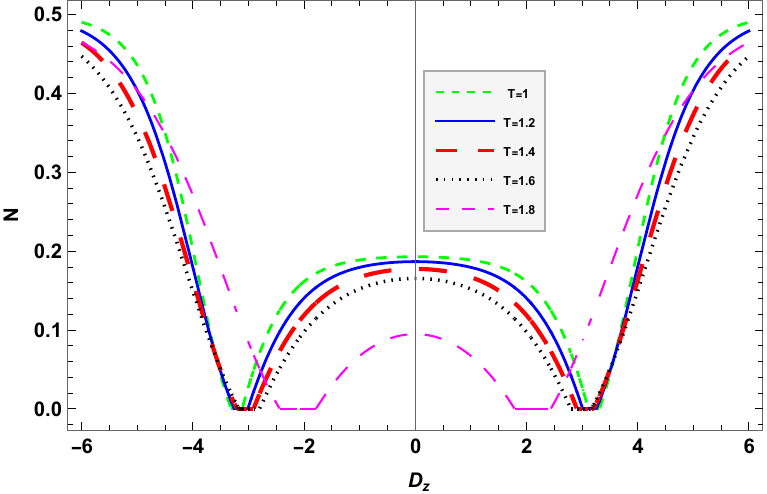} \vfill $\left(d\right)$
 					\end{minipage} \hfill
 					\begin{minipage}[b]{.33\linewidth}
 						\centering
 						\includegraphics[scale=0.45]{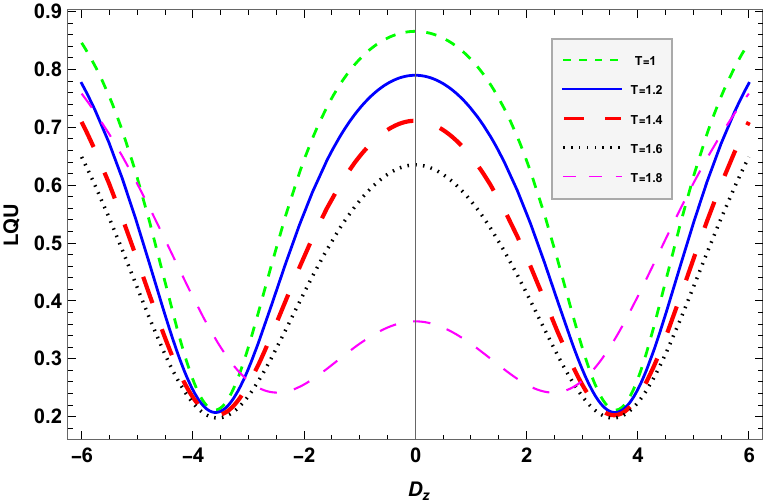} \vfill  $\left(e\right)$
 					\end{minipage} \hfill
 					\begin{minipage}[b]{.33\linewidth}
 						\centering
 						\includegraphics[scale=0.45]{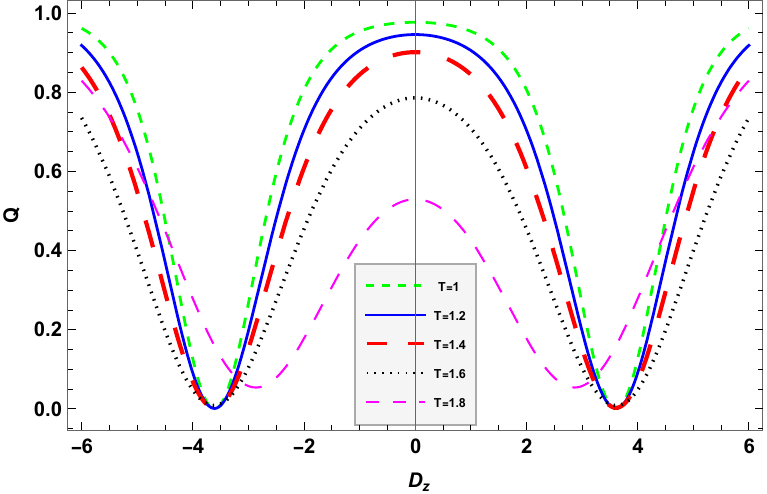} \vfill  $\left(f\right)$
 			\end{minipage}}}
 		 \caption{Negativity $\mathcal{N}$ (a, d), quantum local uncertainty (b, e), and local quantum Fisher information (c, f), versus $D_{z}$ for different values of $T$ in the antiferromagnetic case of $J_{z}=2$ (Top) and  $J_{x}=-1$, $J_{y}=-1.5$, $B=1.5$,  $\Gamma_{z}=0.3$ and ferromagnetic case of $J_{z}=2$ (Bottom).}
    \label{F1}
 		\end{figure}
 		\begin{figure}[hbtp]
 			{{\begin{minipage}[b]{.33\linewidth}
 						\centering
 						\includegraphics[scale=0.45]{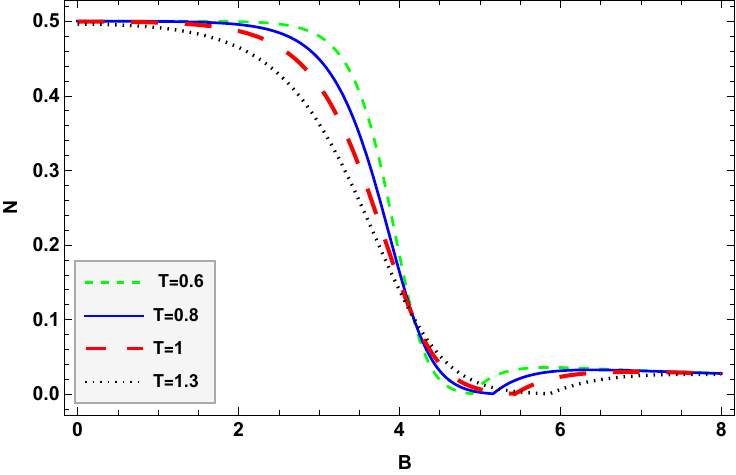} 
 						$\left(a\right)$
 					\end{minipage}\hfill
 					\begin{minipage}[b]{.33\linewidth}
 						\centering
 						\includegraphics[scale=0.45]{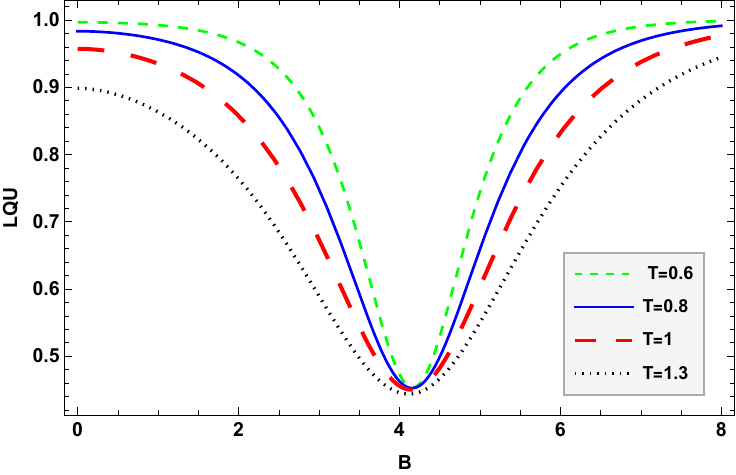} \vfill  $\left(b\right)$
 					\end{minipage}\hfill
 					\begin{minipage}[b]{.33\linewidth}
 						\centering
 						\includegraphics[scale=0.45]{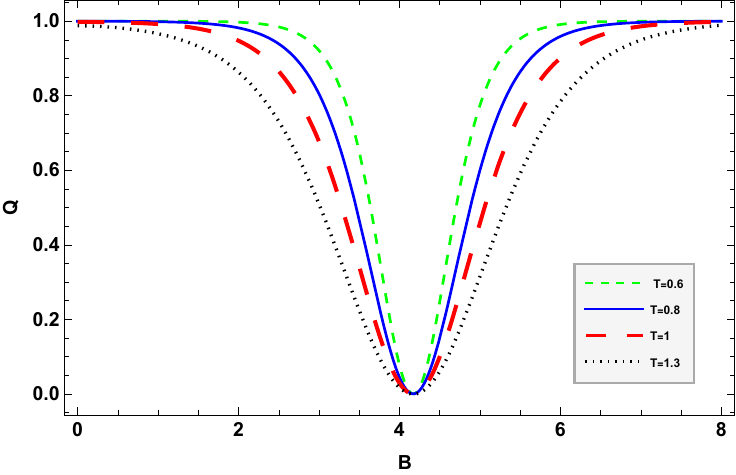} \vfill  $\left(c\right)$
 			\end{minipage}}}
 			\caption{ $\mathbf{a}$ negativity $\mathcal{N}$ ,  $\mathbf{b}$ quantum local uncertainty (LQU) and $\mathbf{c}$  local quantum Fisher information, versus the magnetic field $B$ for different $T$ in the antiferromagnetic case of  $J_{z}=2$ and for $J_{x}=-1$, $J_{y}=-1.5$, $D_{z}=1.8$, $\Gamma_{z}=0.3$.}
 			\label{F22}
 		\end{figure}  
 		\begin{figure}
 			{{\begin{minipage}[b]{.33\linewidth}
 						\centering
 						\includegraphics[scale=0.45]{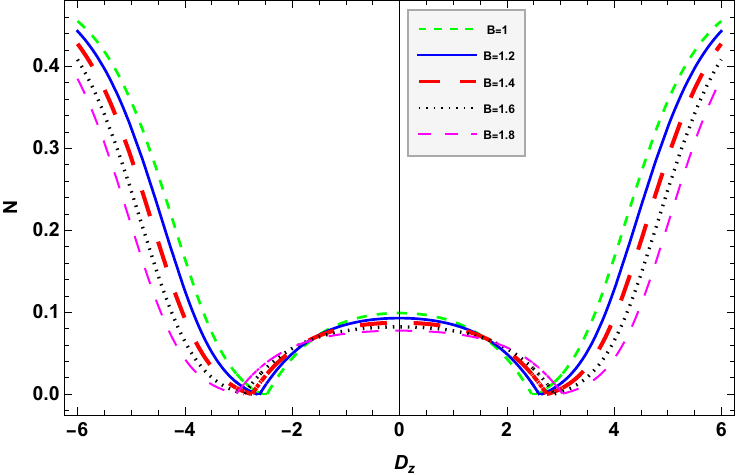}\vfill $\left(a\right)$
 					\end{minipage}\hfill
 					\begin{minipage}[b]{.33\linewidth}
 						\centering
 						\includegraphics[scale=0.45]{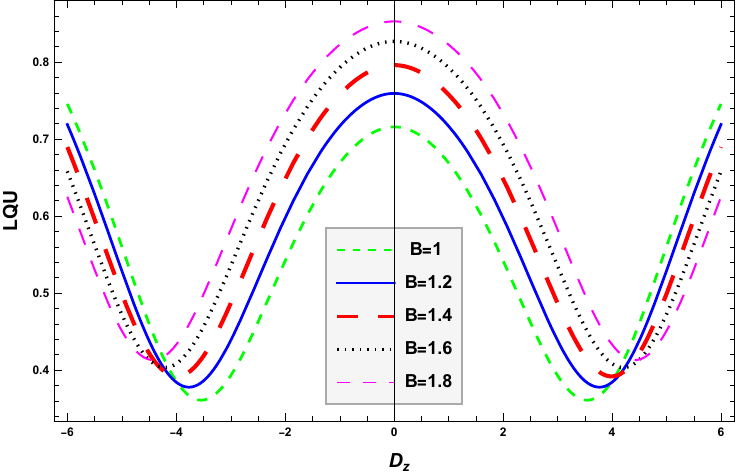} \vfill  $\left(b\right)$
 					\end{minipage}\hfill
 					\begin{minipage}[b]{.33\linewidth}
 						\centering
 						\includegraphics[scale=0.45]{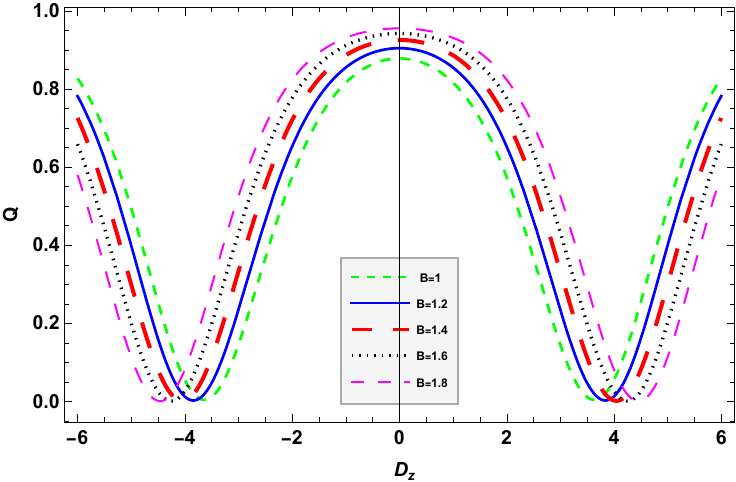} \vfill  $\left(c\right)$
 			\end{minipage}}}
 			\caption{($\mathbf{a}$) negativity $\mathcal{N}$, ($\mathbf{b}$) quantum local uncertainty and ($\mathbf{c}$)  local quantum Fisher information, versus $D_{z}$ for different values of the external magnetic field $B$ in the ferromagnetic   case of $J_{z}=-2$  and for $J_{x}=-1$, $J_{y} =-1.5$, $T=1.5$, and $\Gamma_{z}=0.3$.}
 			\label{F3}
 		\end{figure}
 	\end{widetext}
 	
 	\section{Dynamics Of Negativity, LQU and LQFI under dephasing channel}\label{sec4}
Now, let's delve into investigating the dynamics of non-local correlations under the influence of decoherence effects. This phenomenon presents a significant challenge to the realization of large-scale and low-noise quantum information processing devices. Therefore, the quantum superposition is disrupted by decoherence processes resulting from undesired interactions with the environment \cite{Zurek1981,Zurek1991}. Scientists are particularly interested in understanding how quantum correlations behave when they interact with their environment. A notable discovery, outlined in Ref.\cite{Mazzola2010}, reveals that certain types of quantum correlations, as measured by quantum discord, can exhibit remarkable resistance to decoherence over extended periods. This phenomenon was observed in a system of two qubits subjected to a specific type of environmental influence. The system demonstrated a sudden transition between two states: classical decoherence, where the quantum correlation remains constant while the classical correlation dissipates, and quantum decoherence, wherein the stronger quantum correlation diminishes over time while the classical correlation remains unchanged. In fact, the Markovian evolution of states in noisy environments is typically modeled by quantum channels, which are mappings between operator spaces. Quantum channels, which are completely positive linear maps acting on the quantum state space of a system, play a crucial role in this approach. Mathematically, the Kraus representation provides a comprehensive characterization of the quantum channel's behavior \cite{Bourennane2004,Zurek2003}. 

Let's denote by $\xi_{D}$ the quantum operation that transforms the physical state represented by the input state $\rho_{in}=\varrho$ (i.e., Eq.(\ref{eq14})) into the output state $\rho_{out}=\varrho^{D}(t)$ contingent upon the decoherence probability $\gamma$ as
 	\begin{equation}
 		\varrho^{D}(t)= \xi_{D}(\varrho)=\sum_{i} K_{i} \varrho K_{i}^\dagger,\label{Eq56}
 	\end{equation}
 where the $K_i$ are Kraus operators describing a local coupling and satisfying the completeness relation $K_{i}K^\dagger_{i}=\openone$. Based on the above description, we will examine how Negativity, LQU, and LQFI evolve over time for a two-qubit system undergoing the dephasing channel. Here, we examine a simplified scenario where the quantum channel only affects a single qubit, which we will refer to as qubit $A$. In such a case, we term the channel a one-qubit local dephasing channel and designate it as $\xi_{D}^{A}(\varrho)$ \cite{Yu2003,Chen2022}. The one-qubit dephasing channel $\xi_{D}^{A}(\varrho)$ can be represented by the Kraus operators $K_{1}^{A}$ and $K_{2}^{A}$, defined as follows

\begin{equation}
\begin{split}
K_{1}^{A} &= \sqrt{\gamma}
\begin{pmatrix}
1 & 0 \\
0 & 0
\end{pmatrix} = \frac{\sqrt{\gamma}}{2}\left(\openone_{2\times2} + \sigma_{z}\right), \\
K_{2}^{A} &= \sqrt{\gamma}
\begin{pmatrix}
0 & 0 \\
0 & 1
\end{pmatrix} = \frac{\sqrt{\gamma}}{2}\left(\openone_{2\times2} - \sigma_{z}\right).
\end{split}
\end{equation}
with $K_{i}=K_{i}^{A}\otimes\openone_{2\times2}^{B}$. These Kraus operators and the matrix (\ref{Eq56}) lead to
 	\begin{equation}
 		\varrho^{DC}=(1-\frac{\gamma }{2})\varrho +\frac{\gamma}{2}\sigma_{z} \varrho\sigma_{z},  
 	\end{equation}
 	with $\gamma=1-e^{-\Gamma t}$ and $\Gamma$ indicates the decay rate. After applying the quantum operator $\xi_{D}$, the evolved matrix density $\varrho^{DC}$ under this effect remains of $X$-type. Its expression is as follows
 	\begin{equation}
 		\begin{aligned} 
 			\varrho^{DC}=& \rho_{11}\ket{00}\bra{00}+(1-\gamma)\rho_{22}\ket{01}\bra{01}\\ &+(1-\gamma)\rho_{33}\ket{10}\bra{10}+\rho_{44}\ket{11}\bra{11}\\
 			&+(1-\gamma)|\rho_{14}|[\ket{11}\bra{00}+\ket{00}\bra{11}]\\ 
 			&+(1-\gamma)|\rho_{23}|[\ket{01}\bra{10}+\ket{10}\bra{01}].
 		\end{aligned}\label{Eq59}
 	\end{equation}
  The eigenvalues of the equation (\ref{Eq59}) are given by
 	\begin{align}
 		& \eta_1^{DC} =\frac {e^{\beta J_z}1-\gamma}{Z} (\cosh (\beta r_2)  - \frac{ \xi}{r_2 }),\\
 		& \eta_2^{DC}=\frac{e^{\beta J_z}1-\gamma}{Z} (\cosh (\beta r_2)  + \frac{ \xi}{r_2 }),\\
 		&  \eta_3^{DC}= \frac{ e^{-\beta J_z } }{Z}\left(\cosh (\beta r_3)-\frac{\sinh(\beta r_3)\sqrt{4B^2 +r_1^2(1-\gamma)^2}}{r_3} \right),\\
 		&\eta_4^{DC}=\frac{ e^{-\beta J_z} }{Z}\left(\cosh (\beta r_3)+\frac{\sinh(\beta r_3)\sqrt{4B^2 +r_1^2(1-\gamma)^2}}{r_3} \right),
 	\end{align}
 	with $\xi =\sqrt{ 4 D_z^2\cosh^2( \beta r_2) +(j_x+j_y)^2 \sinh^2 \beta( r_2 )}$, and the corresponding eigenvectors are
 	\begin{align}
 		&\ket{\psi_1^{DC}}=\frac{1}{\sqrt{2}}(0,1,1,0),\hspace{0.5cm}\ket{\psi_2^{DC}}=\frac{1}{\sqrt{2}}(0,-1,1,0),\notag\\
 		&\ket{\psi_3^{DC}}=(\frac{\xi_1}{\sqrt{\zeta_1}},0,0,\frac{1}{\sqrt{\zeta_1}}),\hspace{0.5cm}\ket{\psi_4^{DC}}= (\frac{\xi_2}{\sqrt{\zeta_2}},0,0,\frac{1}{\sqrt{\zeta_2}}),
 	\end{align}
 	where
 	\begin{align}
 		\xi_1=\frac{2B-\sqrt{4B^2+r_1^2(1-\gamma)^2}}{r_1(-1+\gamma)},\hspace{1cm}\zeta_1=\sqrt{\xi_1^2+1},\\
 		\xi_2=\frac{2B-\sqrt{4B^2-r_1^2(1-\gamma)^2}}{r_1(-1+\gamma)},\hspace{1cm}
 		\zeta_2=\sqrt{\xi_2^2+1}.
 	\end{align}
The partial transpose of the density matrix $\varrho^{DC}$ with respect to the first subsystem is
 	\begin{equation}
 		\begin{aligned} 
 			(\varrho^{T_{1}})^{DC}&=\rho_{11}\ket{00}\bra{00}+(1-\gamma)\rho_{22}[\ket{01}\bra{01}+\ket{10}\bra{10}]\\
 			&+\rho_{44} \ket{11}\bra{11} +(1-\gamma)|\rho_{23}| [\ket{11}\bra{00}+ \ket{00}\bra{11}]\\
 			&+(1-\gamma)|\rho_{14}| [\ket{10}\bra{01}+ \ket{01}\bra{10}],
 		\end{aligned}
 	\end{equation}
 and the corresponding eigenvalues are read as
 	\begin{align}
 	e_1^{DC}&=\frac{e^{-\beta J_z}}{Z} \left[\cosh (\beta r_3)-\frac{\sqrt{{\cal P}}}{r_2 r_3 }\right],\\
 	e_2^{DC}&=\frac{e^{-\beta J_z}}{Z} \left[\cosh (\beta r_3)+\frac{{\cal P}}{r_2 r_3 }\right],\\
 	e_3^{DC}&= \frac{1-\gamma}{Z}\left[ e^{\beta J_z} \cosh (\beta  {r_2}) -\frac{e^{-\beta {j_z}} r_1 \sinh (\beta r_3)}{ {r_3} }\right],\\
 	e_4^{DC}&= \frac{1-\gamma}{Z}\left[ e^{\beta J_z} \cosh (\beta  {r_2}) +\frac{e^{-\beta {j_z}} r_1 \sinh (\beta r_3)}{ {r_3} }\right],
 	\end{align}
 	where
 	\begin{align}
 	&\chi= \sqrt{ 4 D_z^2\cosh^2( \beta r_2) +(j_x+j_y)^2 \sinh^2 \beta( r_2 )},\notag\\&{\cal P}=e^{4 \beta J_z}r_3^2 (1-\gamma)^2 \chi +4B^2r_2^2 \sinh ^2(\beta r_3).
 	\end{align}
Whereas, from equation (\ref{eq21}), the expression of the negativity is written as
 	\begin{equation}
 	\mathcal{N}((\varrho^{T_{1}})^{DC}) = \max\{ 0,-2 e^{DC}_{\min} \}.
 	\end{equation} 
Likewise and based on the above formalism for LQU, we find that the matrix ${\cal W}^{DC}=dig\{{\cal W}_{11}^{DC},{\cal W}_{22}^{DC},{\cal W}_{33}^{DC}\}$ where their elements are given by
 	\begin{align}
 		{\cal W}_{11}&=\tau_1+\frac{\xi_+}{4 \tau_1},\hspace{1cm}
 		{\cal W}_{22}=\tau_1+\frac{\xi_-}{4 \tau_1},\\
 		{\cal W}_{33}&=\tau_2+\frac{\zeta_+}{8 \alpha}+\frac{\zeta_-}{8 \beta},
 	\end{align}
 	where
 	\begin{align*}
 		\xi_+ &=\frac{16 r_1 (1-\gamma)^{2} \chi \sinh(\beta r_3)}{ r_2  r_3 Z^2},\\
 		\xi_- &=-\frac{16 r_1 (1-\gamma)^{2} \chi\sinh(\beta r_3)}{ r_2  r_3 Z^2},\\
 		\zeta_+&=\frac{16 e^{-2\beta \text{jz}}\sinh ^2(\beta r_3) \left(4 B^2-r_1^2  (1-\gamma)^2\right)}{r_3^2 Z^2},\\
 		\zeta_-&=\frac{16 e^{-2\beta \text{jz}}\sinh ^2(\beta r_3) \left(4 B^2-r_1^2  (1-\gamma)^2\right)}{r_3^2 Z^2},\\
 		\tau_{1}&=\left(\sqrt{\eta_1^{DC} }+\sqrt{\eta_4^{DC}}\right) \left(\sqrt{\eta_2^{DC}}+\sqrt{\eta_3^{DC}}\right),\\
 		\tau_{2}&=\frac{1}{2}\left(\sqrt{\eta_1^{DC}}+\sqrt{\eta_4^{DC}}\right)^2+\frac{1}{2}\left(\sqrt{\eta_2^{DC}}+\sqrt{\eta_3^{DC}}\right)^{2},\\
 		\alpha&=\left(\sqrt{\eta_1^{DC}}+\sqrt{\eta_4^{DC}}\right)^2,\hspace{.2cm} \beta=\left(\sqrt{\eta_2^{DC}}+\sqrt{\eta_3^{DC}}\right)^2
 	\end{align*}
  Depending on the largest value, the amount of quantum correlation via LQU is yielded by
 		\begin{equation}
 			{\cal LQU}\left(\varrho^{DC} \right)=1-\max\{{\cal W}_{11},{\cal W}_{22} ,{\cal W}_{33}\}.
 		\end{equation}
 For LQFI (Eq.(\ref{eq47})), the elements of the matrix ${\cal M}^{DC}$ are reduced to
 \begin{align}
     {\cal M}_{xx}^{DC}=&2\eta_1^{DC} \left(\frac{  \eta_3^{DC} \frac{1}{  \zeta_1}(\text{$\xi $1}-1)^2}{ \eta_1^{DC}+  \eta_3^{DC}}+\frac{ \eta_4^{DC} \frac{1}{  \zeta_2}(\text{$\xi $2}-1)^2}{ \eta_1^{DC}+\eta_4^{DC}}\right)\notag\\&+ 2  \eta_2^{DC} \left(\frac{  \eta_3^{DC}  \frac{1}{  \zeta_1}(\text{$\xi $1}+1)^2}{ \eta_2^{DC}+  \eta_3^{DC}}+\frac{  \eta_4^{DC}  \frac{1}{  \zeta_1}(\text{$\xi $2}+1)^2}{ \eta_2^{DC}+\eta_4^{DC}}\right),
 \end{align}
 \begin{align}
     {\cal M}_{yy}^{DC}=&2\eta_1^{DC} \left(\frac{  \eta_3^{DC}  \frac{1}{  \zeta_1}(\text{$\xi $1}+1)^2}{ \eta_1^{DC}+  \eta_3^{DC}}+\frac{\eta_4^{DC}  \frac{1}{  \zeta_2}(\text{$\xi $2}+1)^2}{ \eta_1^{DC}+\eta_4^{DC}}\right)\notag\\&+2  \eta_2^{DC} \left(\frac{  \eta_3^{DC}  \frac{1}{  \zeta_1}(1-\text{$\xi $1} )^2}{ \eta_2^{DC}+  \eta_3^{DC}}+\frac{\eta_4^{DC} \frac{1}{  \zeta_2}(1-\text{$\xi $2} )^2}{ \eta_2^{DC}+\eta_4^{DC}}\right),
 \end{align}
 \begin{align}
     {\cal M}_{xy}^{DC}=& -2\eta_1^{DC} i \left(\frac{\left(\text{$\xi $1}^2-1\right)   \eta_3^{DC}}{\zeta_1 ( \eta_1^{DC}+  \eta_3^{DC})}+\frac{\left(\text{$\xi $2}^2-1\right) \eta_4^{DC}}{\zeta ( \eta_1^{DC}+\eta_4^{DC})}\right)\notag\\&+2 i  \eta_2^{DC} \left(\frac{\left(1-\text{$\xi $1}^2\right)   \eta_3^{DC}}{\zeta_1 ( \eta_2^{DC}+  \eta_3^{DC})}+\frac{\left(1-\text{$\xi $2}^2\right) \eta_4^{DC}}{\zeta ( \eta_2^{DC}+\eta_4^{DC})}\right),
 \end{align}
 \begin{align}
     {\cal M}_{yx}^{DC}=&2\eta_1^{DC} i \left(\frac{\left(\text{$\xi $1}^2-1\right)   \eta_3^{DC}}{\zeta_1 ( \eta_1^{DC}+  \eta_3^{DC})}+\frac{\left(\text{$\xi $2}^2-1\right) \eta_4^{DC}}{\zeta_2 ( \eta_1^{DC}+\eta_4^{DC})}\right)\notag\\&+2 i  \eta_2^{DC} \left(\frac{\left(\text{$\xi $1}^2-1\right)   \eta_3^{DC}}{\zeta_1 ( \eta_2^{DC}+  \eta_3^{DC})}+\frac{\left(\text{$\xi $2}^2-1\right) \eta_4^{DC}}{\zeta_2 ( \eta_2^{DC}+\eta_4^{DC})}\right),
 \end{align}
 \begin{align}
 {\cal M}_{zz}^{DC}=&\frac{4  \eta_1^{DC}  \eta_2^{DC}}{ \eta_1^{DC}+ \eta_2^{DC}}+\frac{4   \eta_3^{DC} \eta_4^{DC}(\text{$\xi $1}\text{$\xi $2} -1)^2}{\zeta_{1}\zeta_{2}(\eta_3^{DC}+\eta_4^{DC})}.
\end{align}
   It is easy to check that the eigenvalues of the matrix ${\cal M}^{DC}$ are equal to $\lambda_{3}^{DC}= {\cal M}_{zz}^{DC}$ and
   \begin{align}
       \lambda_{1}^{DC}=\frac{-1}{2}&\left[\sqrt{({\cal M}_{xx}^{DC})^2-2 {\cal M}_{xx}^{DC} {\cal M}_{yy}^{DC}+4 {\cal M}_{xy}^{DC} {\cal M}_{yx}^{DC}+({\cal M}_{yy}^{DC})^2}\right.\notag\\&\left.-{\cal M}_{xx}^{DC}-{\cal M}_{yy}^{DC}\right],
   \end{align}
   \begin{align}
       \lambda_{2}^{DC}=\frac{1}{2}&\left[\sqrt{({\cal M}_{xx}^{DC})^2-2 {\cal M}_{xx}^{DC} {\cal M}_{yy}^{DC}+4{\cal M}_{xy}^{DC} {\cal M}_{yx}^{DC}+({\cal M}_{yy}^{DC})^2}\right.\notag\\&\left.+{\cal M}_{xx}^{DC}+{\cal M}_{yy}^{DC}\right].
   \end{align}
 It turns out that $\lambda_2^{DC}\geq\lambda_1^{DC}$. Hence, the analytical expression of  $\mathcal{Q}$ is written in the form
 		\begin{equation}
 			\mathcal{Q}\left(\varrho^{DC} \right) =1-\max\{ \lambda_2^{DC},\lambda_3^{DC}\}.
 		\end{equation}
   
 \begin{widetext}
 
	\begin{figure}[hbtp]
   {{\begin{minipage}[b]{.3\linewidth}
 						\centering
 						\includegraphics[scale=0.45]{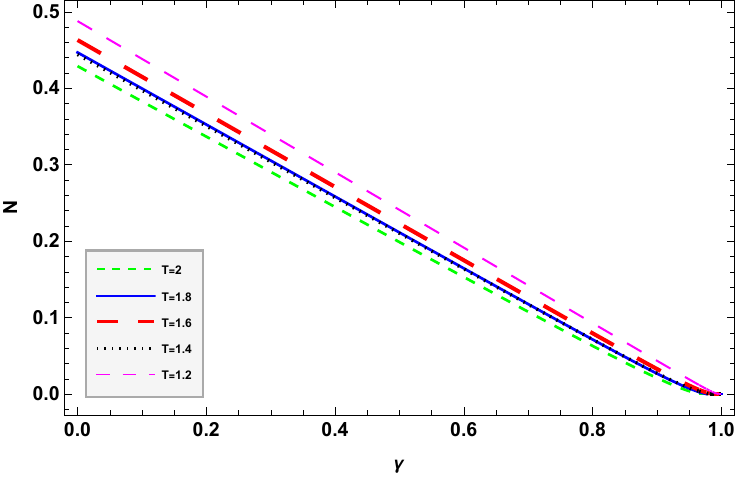} \vfill $\left(a\right)$
 					\end{minipage} \hfill
 					\begin{minipage}[b]{.3\linewidth}
 						\centering
 						\includegraphics[scale=0.45]{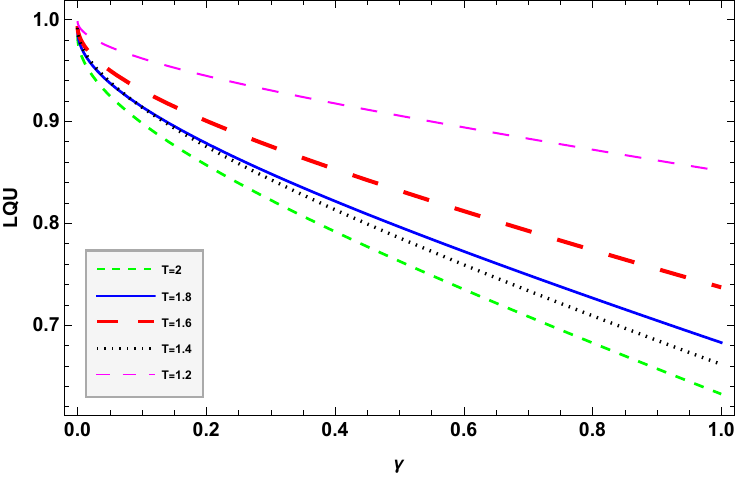} \vfill  $\left(b\right)$
 					\end{minipage} \hfill
 					\begin{minipage}[b]{.3\linewidth}
 						\centering
 						\includegraphics[scale=0.45]{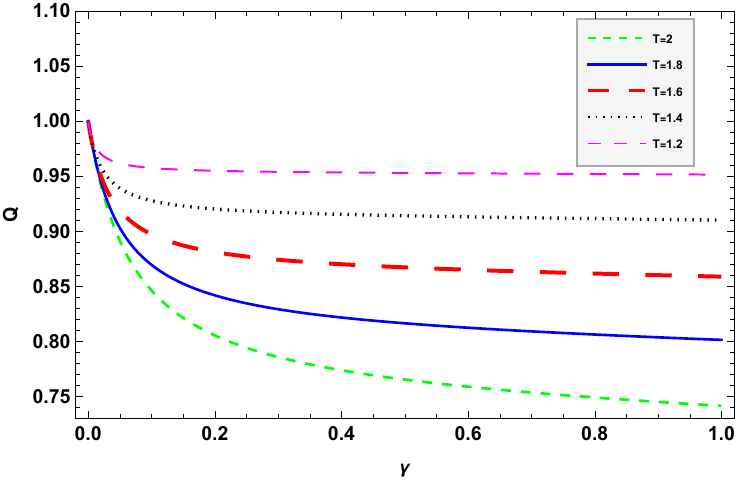} \vfill  $\left(c\right)$
 			\end{minipage}}}
  {{\begin{minipage}[b]{.3\linewidth}
 						\centering
 						\includegraphics[scale=0.45]{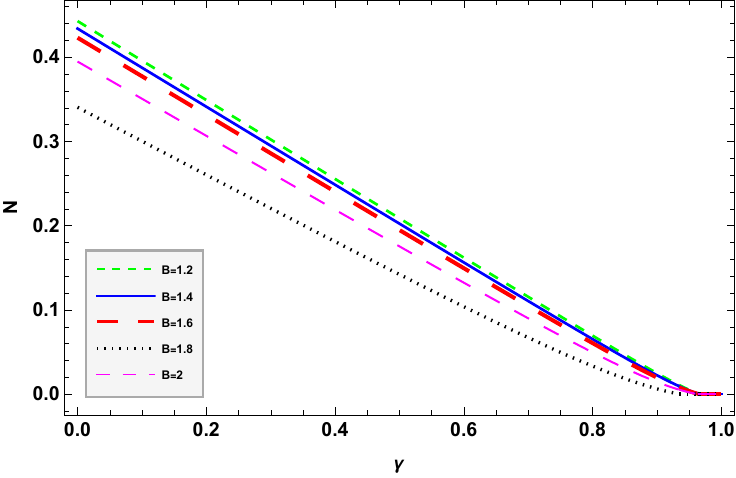} \vfill $\left(d\right)$
 					\end{minipage} \hfill
 					\begin{minipage}[b]{.3\linewidth}
 						\centering
 						\includegraphics[scale=0.45]{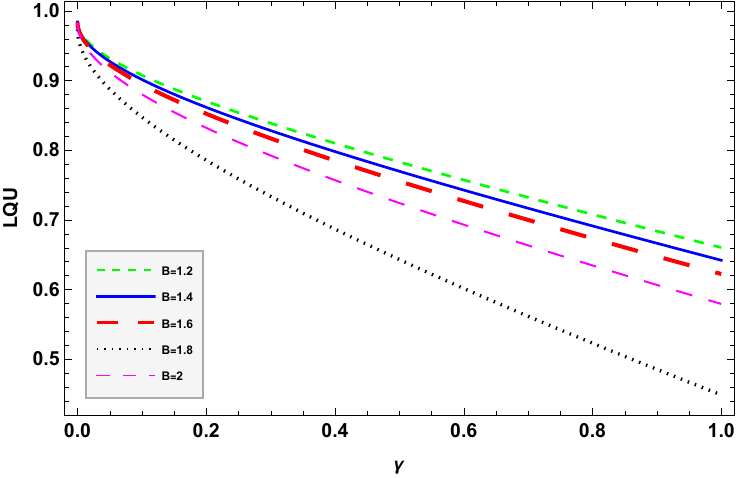} \vfill  $\left(e\right)$
 					\end{minipage} \hfill
 					\begin{minipage}[b]{.3\linewidth}
 						\centering
 						\includegraphics[scale=0.45]{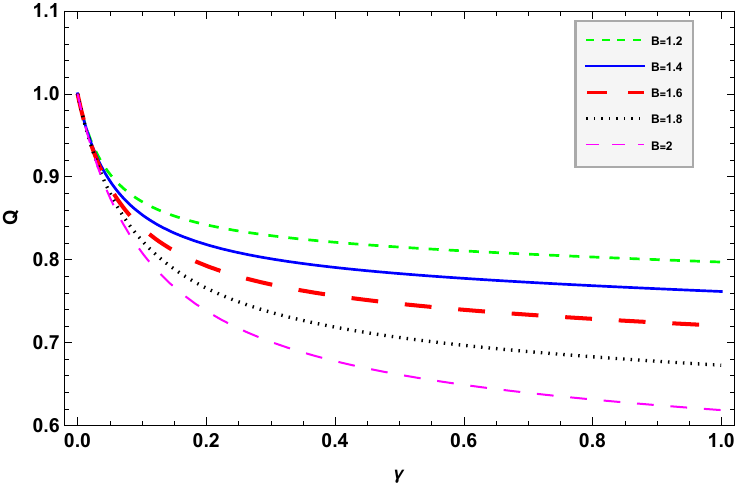} \vfill  $\left(f\right)$
 			\end{minipage}}}
 		\caption{Variation of Negativity (panels $a$ and $d$), LQU (panels $b$ and $e$), and LQFI (panels $c$ and $f$) with respect to the decoherence parameter $\gamma$ under dephasing channel. The variations are shown for both various temperature $T$ with $B$ fixed (top row) and various magnetic field $B$ with $T$ fixed (bottom row).}\label{F4}
 \end{figure}
 \end{widetext}
Figure (\ref{F4}) illustrates how Negativity, LQU, and LQFI behave with increasing decoherence parameter $\gamma$, providing key insights into quantum correlations dynamics under noise channels. As shown in panel ($a$), the peak negativity at $\gamma=0$ suggests that minimal decoherence preserves quantum entanglement. Additionally, consistent with theoretical expectations, correlation increases as temperature decreases. Conversely, at $\gamma=1$, excessive decoherence eliminates entanglement, highlighting its detrimental effect. Therefore, the amount of negativity strongly depends on the decoherence parameter $\gamma$.\par

Now let's analyze the influence of decoherence on local quantum uncertainty. Figure~\ref{F4}($b$) depicts the behavior of LQU dynamics versus the decoherence parameter across different temperatures. In panel ($e$), we extend this analysis to various magnetic field strengths. Interestingly, weaker magnetic fields appear to enhance non-classical correlations, suggesting a more favorable environment for maintaining these correlations. This phenomenon likely arises from reduced external disturbances that could disrupt quantum coherence. Notably, the highest correlation occurs once again at $\gamma=0$, emphasizing the critical role of minimizing decoherence to preserve quantum correlations under varying environmental conditions. As expected, the results reported in Figure~\ref{F4}($b$) and ($e$) confirm that for large values of temperature $T$ and magnetic field $B$, quantum correlations tend to be destroyed.\par

Figure \ref{F4}($c$)-($f$) display the local quantum Fisher information versus the decoherence parameter. As seen in panel ($c$), quantum correlations decrease with increasing $\gamma$ for different temperatures. Additionally, higher temperatures accelerate decoherence. Panel ($f$) shows similar results for the magnetic field. We can therefore conclude that both temperature and magnetic field influence the amount of non-classical correlations. This is likely due to interactions with the environment that affect the superposition of states, leading to decoherence. The key result from these figures is that the amount of quantum correlation depends heavily on the value of decoherence parameter, reaching its maximum at $\gamma\rightarrow 1$ and its minimum at $\gamma\rightarrow 1$. Furthermore, comparing the findings in Figure (\ref{F4}) for the dephasing channel, we observe a strong decrease in negativity, while Q and LQU exhibit similar behavior. Interesting, LQFI shows surprising resilience compared to negativity and LQU, even exhibiting a temporary freezing effect.
 
 \section{Concluding remarks}\label{sec5}
We investigated the dynamics of three quantum correlation measures (Negativity, LQU, and LQFI) in a two-qubit, fully anisotropic Heisenberg model under the influence of Dzyaloshinsky–Moriya (DM) and Kaplan–Shekhtman–Entin-Wohlman–Aharon (KSEA) interactions, along with a magnetic field. All three quantifiers exhibited significant agreement. DM and KSEA interactions strongly affected non-classical correlations \cite{Park2019}.  In the ferromagnetic regime, DM coupling led to a minimum in correlations. A magnetic field, combined with the KSEA interaction, could create both minima and maxima in nonlocal correlations. Notably, the antiferromagnetic and ferromagnetic regimes displayed distinct behaviors. Correlations increased with the DM parameter $D_z$ in the antiferromagnetic case. In contrast, the ferromagnetic case exhibited a more intricate pattern, with correlations vanishing in some regions before reappearing and reaching a maximum. This vanishing phenomenon can be attributed to the interplay of DM and KSEA interactions.\par
 
Interestingly, all three measures reach a maximum as $D_{z}$ increases, suggesting the states approach maximal correlation. This highlights the role of these interactions in driving quantum correlation dynamics. We further investigated the same measures under decoherence using the same model with a dephasing channel. As expected, entanglement is suppressed by decoherence. Interestingly, under the dephasing channel, these three measures (negativity, LQU, and LQFI) exhibit different behavior. Notably, LQU and LQFI perform identically, but LQFI demonstrates surprising resilience to decoherence. In some instances, it even exhibits a temporary freezing effect. This result provides valuable physical insights and suggests the potential for importance implementation of the LQFI measure in noisy environments.

\section*{ACKNOWLEDGMENTS}
S.G. acknowledges the financial support of the National Center for Scientific and Technical Research (CNRST) through the “PhD-Associate Scholarship-PASS” program. The authors are deeply indebted to the reviewers for their thoughtful comments, which significantly enhanced the quality of this paper.

 \end{document}